# Effects of opposite atoms on electronic structure and optical absorption of two-dimensional hexagonal boron nitride

You-Zhao Lan*[1]

*Key Laboratory of the Ministry of Education for Advanced Catalysis Materials, College of Chemistry and Life Sciences, Zhejiang Normal University, Jinhua, Zhejiang, 321004, China*

**Abstract**

We perform the first-principles many-body GW and Bethe-Salpeter equation (BSE) calculations on the two-dimensional hexagonal boron nitride (2D-*h*BN) to explore the effects of opposite atoms on the electronic structure and linear one-photon absorption (OPA). Five AA- and AB-stacked bilayer and eight AAB-stacked trilayer structures are considered. The AAB-stacked trilayer *h*BN (TL-BN) structures are constructed by mixing the AA- and AB-stacked bilayer *h*BN (BL-BN). We show that the GW approximation gives rise to different types (*i.e.*, indirect or direct) of fundamental band gaps from the independent particle approximation for all structures except those dominated by the B–B opposite. The stacking modes dominated by the B–B opposite have a direct fundamental band gap in both approximations. The OPA spectra are calculated by solving the Bethe-Salpeter equation combined with the GW quasi-particle correction. Strong absorption peaks are found for most structures in the deep-ultraviolet region. The binding energy and Davydov splitting of excitons of TL-BN strongly depend on the opposite atoms and are related to the role of the stacking BL-BN substructure. Finally, taking the six-layer and below AB-stacked structures as examples, we show that the B–B opposite unit is helpful in constructing the turbostratic-phase-like stacking structures with a direct fundamental band gap which are more suitable for optoelectronic applications.

1. **Introduction**

The electronic structure and optical property of the two-dimensional hexagonal boron nitride (2D-*h*BN) have attracted much attention [1–13] in recent years. Experimental and theoretical studies have consistently shown that 2D-*h*BN has a wide bandgap, but there are contradictions on the size and type of fundamental band gap (FBG) (*i.e.*, indirect or direct), even for the monolayer BN (ML-BN) that has been widely studied [7,14–16]. For ML-BN, the density functional theory (DFT) calculations predict the FBGs with a range of 4.2 – 4.7 eV between the valence band maximum (VBM) at the K point and the conduction band minimum (CBM) at various points (K, M, or Γ point) [7,14,15]. The GW approximation (GWA) calculations dramatically increase the energy gap by ~ 2.5 eV and also lead to a change of FBG from direct to indirect [7,15]. A recent experimental study confirmed a direct FBG of ~ 6.1 eV for ML-BN by using the reflectance and photoluminescence experiments in the deep-ultraviolet region [2]. Similarly, for the bulk $h$BN, the energy band structure strongly depends on the stacking mode [12,17,18]. The FBG ranges from 3.1 to 4.5 eV based on the DFT/PBE and DFT/LDA calculations [17,18]. Direct and indirect FBGs are also theoretically found in different stacking modes, similar to the experimental studies [19,20].

---
[1] Corresponding author: lyzhao@zjnu.cn

For few-layer $h$BN (FL-BN) structures, which have more adjustable parameters, such as stacking mode and layer number, they have more changeable electronic structure properties than ML-BN and bulk $h$BN [4,7,13,21–23]. For example, for the AA′- and AB-stacked structures, the FBG changes from direct to indirect as the structure changes from ML-BN to FL-BN [13]. The AA′-stacked bilayer has distinctly different excitonic response from the AB-stacked one [4]. As an increase of the number of layers, the emission peaks exhibit a monotonic blue-shift [21]. The measurements of optical second harmonic generation show strong enhancement in the AB-stacked structure relative to monolayer and AA′-stacked bilayer, though similar linear absorption spectra were measured for these three structures [22]. Theoretical calculations [7] on the AA′-stacked FL-BNs up to five layers show that the excitonic spectra are resolved by surface and inner excitons and interesting Davydov splitting. Under biaxial strain, the size and type of FBG of BL-BN are tunable [24], which indicates possibly various applications in electronic and optoelectronic devices.

Note that most studies focus on the singly ordered stacking modes, such as AA′ and AB stackings, while the randomly stacked and disordered phase, namely turbostratic ($t$-BN) phase, was found in experiments sixty years ago [25–27]. Recently, Mengle and Kioupakisa [23] studied five $t$-BN structures containing ten randomly chosen layers and found that the $t$-BN structures had a quasi-direct FBG which only allows weakly direct optical transitions. The random stacking breaks the symmetry of originally ordered FL-BN, which further results in changes in electronic structures and optical properties. Obviously, the random stacking combined with a change of the layer number will lead to a large number of FL-BNs. In particular, there can be a great variety of assignments of opposite atoms in these structures, which leads to some new unknown structure-property relations. Meanwhile, as mentioned above, FL-BN can have direct or indirect FBG depending on the stacking mode. Different stacking modes can lead to different opposite atoms, then to different interlayer interactions, and ultimately to different energy bands with direct or indirect FBG. It is beneficial to filter out the direct-gap FL-BN because materials with direct FBG have higher optical efficiency in optoelectronic applications, such as light emitting diodes and semiconductor laser.

In this work, to explore the effect of opposite atoms on properties, we select five bilayer BN structures (*i.e.*, two AA and three AB stackings) and eight AAB-stacked trilayer BN (TL-BN) structures formed by mixing AA and AB stacking modes, and calculate their electronic structures and linear optical properties. All possible atomic opposites are considered in these stackings. The electronic structures are calculated within both the independent particle approximation (IPA) and the GWA. The linear one-photon absorption (OPA) spectra are calculated by solving the Bethe-Salpeter equation (BSE) combined with the GWA quasi-particle correction. Based on the results of BL- and TL-BN structures, we further explore the *t*-BN-like randomly AB-stacked six-layer structures to filter out the structures with a direct FBG.

In Section 2, we describe the computational details including geometry and GW+BSE calculations. In Section 3, we discuss the calculated electronic structures and OPA and construct the FL-BN structures which have a direct FBG based on the B–B opposite substructure. Conclusions are given in Section 4.

## 2. Computational methods

### 2.1 *Geometry*

The two-dimensional bilayer (BL-) and trilayer (TL-) hexagonal BN structures with different stacking modes are shown in Fig. 1. We consider AA and AB-stacked bilayer structures and their mixture to form the trilayer structures. The AB-stacked structures mean two layers overlap with one set of atoms facing each other. In terms of different sets of opposite atoms, there are three AB stackings labeled by AB-NN, AB-BN, and AB-BB. The AA-stacked structures mean two layers overlap with all atoms facing each other. We constructed two AA stackings labeled by AA-NN and AA-BN. Note that AA-NN can be also called by AA-BB. For mixture of AA and AB stackings, we constructed eight structures in terms of different opposite atoms, *i.e.*, AAB-BNN, AAB-NBB, AAB-BNB, AAB-BBB, AAB-NNN, AAB-NNB, AAB-BBN, and AAB-NBN. The definition of labels is given in the figure caption. Hereafter, for simplicity, we call trilayer structures by opposite atoms only, namely BNN, NBB, BNB, etc.

The initial structures were constructed based on the bulk *h*BN structure [28]. As a reference, we optimized the bulk *h*BN by using the same method and obtained the structure parameters ($a = b = 2.503$ Å, $c = 6.681$Å) which are in agreement with the experimental parameters [28] ($a = b = 2.498$Å, $c = 6.636$Å). All the structures were optimized using the DFT within the GGA-PBE approximation combined with the pseudopotential plane wave method, as implemented in the PWSCF code.[29] A k-point mesh of 6×6×1, a force threshold of 0.01 eV/Å, and a stress threshold of 0.02 GPa were used for the optimizations. The relaxation of the unit cell was included in optimizations. The optimized lattice parameters are shown in Fig. 1. A vacuum spacing larger than 13 Å was used to ensure negligible interaction between the slabs. The van der Waals interaction was taken into account by using the Tkatchenko-Scheffler (TS) dispersion corrections [30]. The interlayer distances and cell parameters of all optimized structures are given as inset tables in Fig. 1.

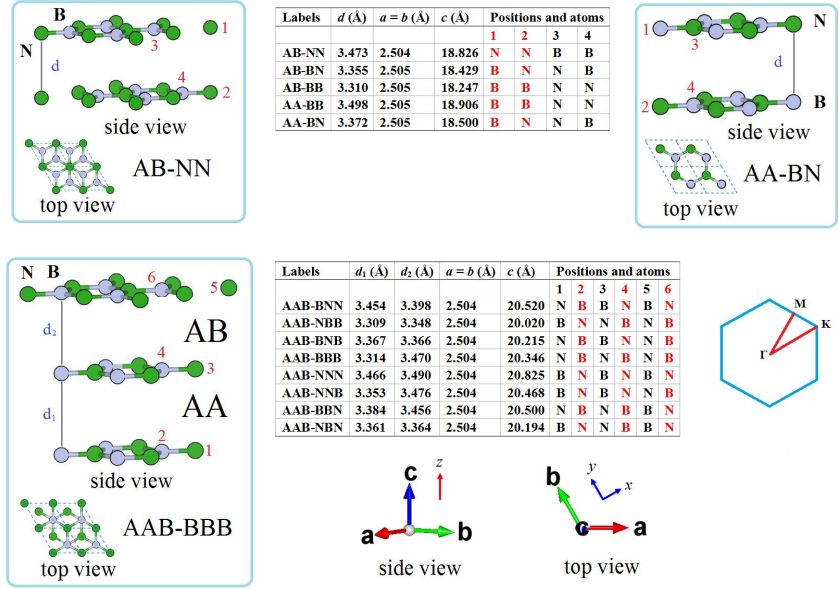

Figure 1. *The optimized structures (side and top views) of BL- and TL-BN. The labels of BL-BN are defined by AB-XY and AA-XY, where X and Y are the opposite atoms at positions 1 and 2, respectively. For example, AB-NN and AA-BN are shown explicitly. The labels of TL-BN are similarly defined by AAB-XYZ, where X, Y, and Z are opposite atoms at positions 2, 4, and 6, respectively. For example, AAB-BBB is shown explicitly. The inset tables list the interlayer distances, cell parameters, and atoms at selected positions of all optimized structures.*

2.2 *GW+BSE calculations*

Since the DFT within the GGA-PBE approximation usually underestimates the energy gap of materials, we performed the many-body GWA (one-shot level or $G_0W_0$) calculations to correct the energy bands. The GWA calculations were based on the plasmon pole approximation as implemented in the Yambo package [31] which reads the band structures and wave functions of ground state calculated within the IPA as implemented in the PWSCF code.[29] The GGA-PBE combined with the pseudopotential plane wave method was used to calculate the ground state. The convergence tests were performed on the ML-BN which has been widely studied [3,5,11,13,32–34]. Three parameters, namely k-grid, response block size in polarizability matrix, and the number of empty states in dielectric function, were considered. As shown in table S1, we obtain the converged $G_0W_0$ energy gaps for both an indirect energy gap of 6.57 eV between the K point and the Γ point and a direct energy gap of 7.24 eV at the K point, in agreement with previous calculations [3,13,32]. The corresponding three parameters are 30×30×1, 10 Ry, and 200 empty states, respectively. Finally, for BL-BN and TL-BN, we used the 30×30×1, 10 Ry, and 300 empty states in the $G_0W_0$ calculations, which produced the $G_0W_0$ gaps within an accuracy of ~0.03 eV.

We calculated the optical spectra based on the solution of the BSE:[35]

$$\left(E_{ck}-E_{vk}\right)A^S_{vck} + \sum_{k'v'c'} \left\langle vck|K_{eh}|v'c'k'\right\rangle A^S_{v'c'k'} = \Omega^S A^S_{vck} \quad (1)$$

The excited state $S$ is given by the linear combination of independent-particle excitations |vck> (i.e., valence band |vk> to conduction band |ck>) as

$$|S\rangle = \sum_{c,v,k} A^S_{c,v,k} |vck\rangle \quad (2)$$

The interaction kernel $K_{eh}$ includes the screened Coulomb interaction between electrons and holes, and the exchange interaction, which includes the so-called local-field effect. When the $K_{eh}$ is ignored, Eq.1 reduces to independent particle excitations. We used the Coulomb cutoff technique [36–38] and the corresponding length cutoff was set to a slightly smaller value than the $c$ lattice parameter (Fig. 1) of supercell. In this case, we prefer to use the imaginary part of two-dimensional polarizability to understand the optical absorption of materials [36]. The two-dimensional polarizability is defined by [37,39]:

$$\chi^{2D} = L\frac{\varepsilon-1}{4\pi} \quad (3)$$

, where $L$ is the effective thickness, which is assumed to $c$ lattice parameter (Fig. 1), and $\varepsilon$ is the dielectric constant. The imaginary part ($\varepsilon_2$) of $\varepsilon$ can be understood by [36]:

$$\varepsilon_2(\omega) = \frac{8\pi^2}{q^2}\lim_{q\to 0}\sum_S \left|\sum_{c,v,k} A^S_{vck}\left\langle v(k-q)|e^{-iq\cdot r}|ck\right\rangle\right|^2 \delta\left(\Omega^S-\omega-\eta\right) \quad (4)$$

, where $\eta$ is the damping factor and set to 0.1 eV.

For the BSE calculation, we also carried out the convergence tests on the k-grid and the IPA bands used to construct the electron-hole basis (*eh*-basis) of the BSE kernel ($K_{eh}$). Other parameters (*i.e.*, response block size in polarizability matrix and the number of empty states in dielectric function) are the same as those used in the $G_0W_0$ calculations. Since the valence and conduction band dispersions based on the IPA are somewhat different from those based on the GWA (see Fig. 2 below), we used a quasi-particle correction for the entire energy band, rather than a simple scissor correction. As an example, convergence tests on k-grids and *eh*-basis were performed on ML-BN, while convergence tests on *eh*-basis were performed on AA-BN and BNN, and corresponding results are shown in Fig. S1. A k-grid of 30×30×1 leads to a good convergence for the first and second absorption peaks. For the fixed 30×30×1 k-grid, the highest four valence bands and the lowest four conduction bands are enough to obtain the converged first two (see 1–8 of ML-BN and 5–12 of AA-BN) or three (see 9–16 of BNN) absorption peaks. This is due to that these absorption peaks mainly arise from the transitions between the valence and conduction bands near the Fermi level (see below). Finally, we adopt the 30×30×1 k-grid for all BSE calculations and the *eh*-basis of 5–12 and 9–16 for BL-BN and TL-BN, respectively.

## 3. Results and discussion

### 3.1 *Band structures*

Figure 2 shows the band structures of BL-BN and TL-BN based on the IPA and GWA calculations. For comparison, the band structure of ML-BN is included. First, from IPA to GWA, the type of the FBG changes significantly. For ML-BN (Fig. 2a), the IPA yields a direct FBG at K point, while the GWA yields an indirect one between K and Γ points, in agreement with previous reports [7,14–16]. There are two possible reasons for inconsistency in the type of FBG between IPA and GWA. One is that the lowest unoccupied conduction band (LUCB) along the M-K path is very flat and the conduction band bottoms (CBBs) at Γ and K points have only a small energy difference of 20 meV. The other pointed by Blase *et al.*[32] is that the self-energy correction in ML-BN is strongly k-point dependent and has more effect on the K and M points than the Γ point. This inconsistency also exists in the band structures of BL-BN and TL-BN. For example, for AB-BN (Fig. 2d), the IPA yields very close K – K (4.59 eV) and K – M (4.54 eV) gaps, while the GWA results in an indirect K – Γ gap of 6.31 eV. A similar case occurs in the band structure of NNB (Fig. 2m) which has very close K – K (4.05 eV) and K – M (4.09 eV) gaps within the IPA and has an indirect K – Γ gap of 5.50 eV within the GWA. Note that for five BL-BN structures considered here, Mengle and Kioupakis [23] has also calculated the band structures by using the $G_0W_0$ method. Comparing our results with theirs, we observe qualitative consistency but some quantitative differences due to differences in computational parameters, such as k-grid and vacuum spacing in supercell. Overall, the GWA does not change the relative magnitude of the CBBs at M and K points, but changes the CBB at Γ point relative to those at M and K points.

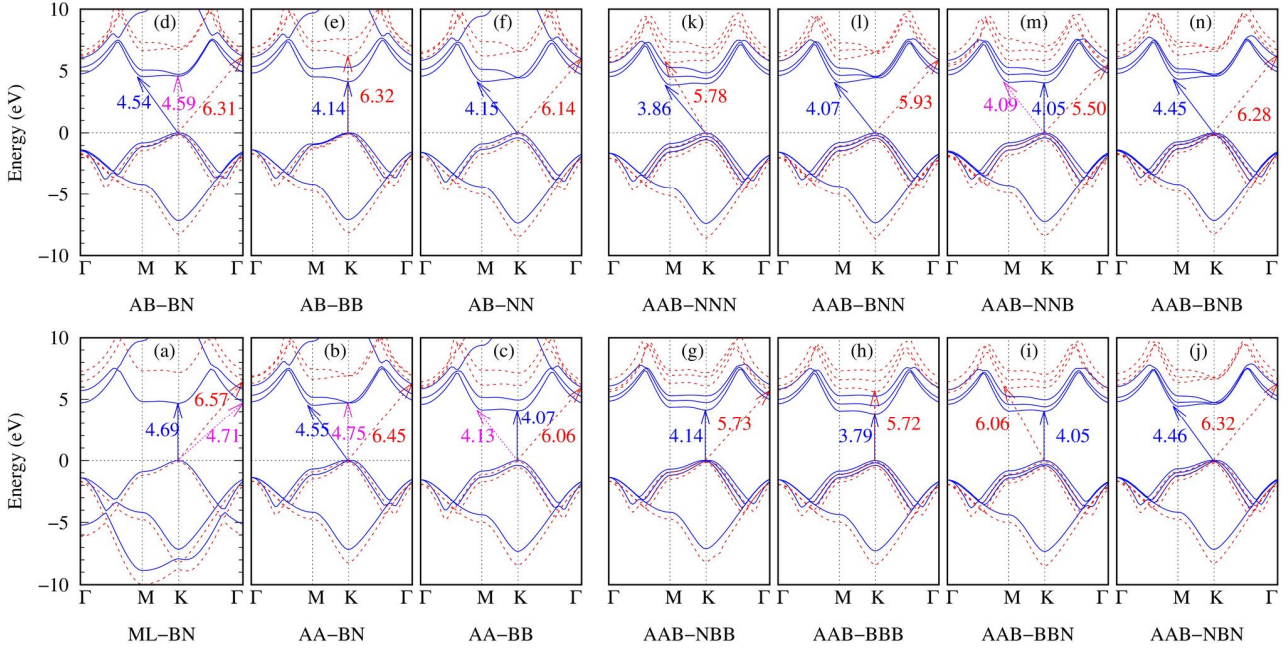

Figure 2. *Band structures of ML-, BL-, and TL-BN based on the IPA (blue solid line) and GWA (red dash line) calculations. The energy gaps between the conduction band minimum and the valence band maximum are shown by arrows and values. The valence band maximum is set to zero.*

Second, for BL-BN, we observe that the stackings with only B–B (AB-BB) and N–N (AB-NN) opposites separately have a direct and indirect FBG, and that those with B–N (AB-BN and AA-BN) or with both B–B and N–N (AA-BB) have an indirect FBG. The AB-BB (Fig. 2e) with only B–B opposite has direct FBGs of 4.14 and 6.32 eV within the IPA and GWA, respectively. In the band structure of AB-BB, the M-K path shows a strong dispersion. The AB-NN with only N–N opposite (Fig. 2f) has an indirect FBG within the IPA and GWA. The AA-BB with both the B–B and N–N opposites (Fig. 2c) has a direct FBG within the IPA but an indirect one within the GWA. The AB-BN with only B–N opposite (Fig. 2d) has an indirect FBG within the IPA and GWA. Similar to the ML-BN (Fig. 2a), the AA-BB and AB-BN also have a relatively flat M-K path (e.g., see gaps of 4.54 and 4.59 eV in Fig. 2d) and a close CBB energy at K and Γ points. For ease of understanding, Fig. 3a shows the scheme of dependence of M-K path on the opposite atoms. As shown in Fig. 2, for BL-BN, the highest occupied valence band (HOVB) of all five structures exhibit a similar M-K path with a higher energy at K point than M point, which is illustrated as the bottom line in Fig. 3a. The valence band maximum (VBM) locates at K point. However, the dispersion of the LUCB along M-K path strongly depends on the opposite atoms. The N–N and B–B opposites separately lead to a dispersion with a higher (red line in Fig. 3a) and lower (blue line in Fig. 3a) energy at K point than M point; thus the AB-BB (Fig. 2e) has a direct FBG while the AB-NN (Fig. 2f) has an indirect one. For B–N or B–B + N–N, the mixture of B–N interactions results in a relatively flat dispersion (yellow and green lines in Fig. 3a).

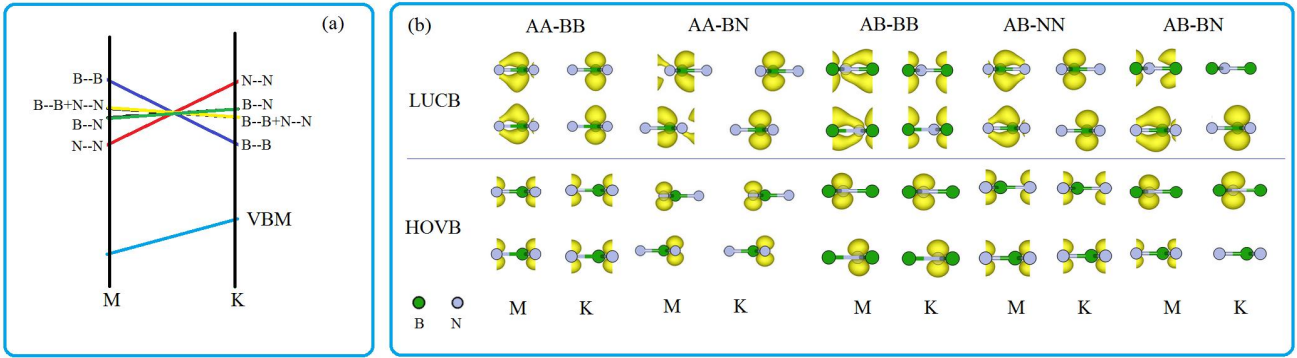

Figure 3. *(a) Scheme of dependence of M-K path on the opposite atoms. (b) Charge density distributions of HOVB and LUCB at M and K points for BL-BN.*

To elucidate the different dispersions of M-K path, we examine the charge density distributions (Fig. 3b) of HOVB and LUCB at M and K points for all five BL-BN structures. As shown in Fig. 3b, the HOVBs of all BL-BNs are derived from the $p_z$ state of N atom and hardly from that of B atom, and thus the HOVBs of all BL-BNs have a similar dispersion (Fig. 2). In contrast, the LUCBs of all BL-BNs are very different and strongly depend on the opposite atoms. The LUCB of AA-BB, with both B–B and N–N opposites, is more like that of AB-BB than AB-NN, that is, LUCBs of AA-BB and AB-BB at M and K points are derived from the $p_z$ state of *directly* opposite B atoms while the LUCB of AB-NN at K point are derive from the $p_z$ state of *diagonally* opposite B atoms. Thus, the M-K path of AA-BB is relative flat but trends to that of AB-BB with a higher energy at M point than K point [see blue and yellow lines in Fig. 3a and gaps of 4.13 (K–M) and 4.07 (K–K) eV in Fig. 2c]. The LUCB of AA-BN is more like that of AB-NN than AB-BB, and thus the M-K path of AA-BN is relative flat but trends to that of AB-NN with a lower energy at M point than K point [see red and green lines in Fig. 3a and gaps of 4.55 (K–M) and 4.75 (K–K) eV in Fig. 2b]. Finally, the M-K path of AB-BN is relative flat but trends to that of AB-NN [see gaps of 4.54 (K–M) and 4.59 (K–K) eV in Fig. 2d] because the lower layer of LUCB of AB-BN is almost the same as that of AB-NN.

Finally, for TL-BN, the band structures of all eight structures are shown in Figs. 2(g–n). Overall, the band structures of TL-BN have very similar characteristics to those of BL-BN. All the TL-BN structures have similar valence band dispersions, and thus the difference in FBG depends on the conduction band dispersion. Since the VBM of all TL-BN structures locates at K point, the FBG will be determined by the CBB at M, K, or Γ point. Meanwhile, the type of FBG of TL-BN also dramatically depends on the opposite atoms. We note that the effect of the mixture of AA and AB stackings on the type of FBG. The substacking with B–B opposite is helpful in forming the direct FBG. For example, NBB formed by mixing the AB-BB and AA-BN has a direct FBG (Fig. 2g). AB-BB has a direct FBG with CBM at K point (Fig. 2e), while AA-BN has an indirect FBG with CBM at M point (Fig. 2b), which leads to a direct FBG at K point for NBB. Again, BNB (Fig. 2n) is formed by mixing AA-BN and AB-BN. Since both substackings have an indirect FBG (see Fig. 2b and Fig. 2d), BNB ultimately has an indirect FBG. The BBB structure with two B–B opposites has a direct FBG. Moreover, for structures with the largest number of B–B opposites, the GWA correction does not change the type of FBG (Fig. 2e and Fig. 2h). For other structures, the GWA corrections mostly change the type of FBG or the position of CBM. In subsection 3.3, based on the B–B opposite substacking, we will further discuss the construction of the FL-BN with a direct FBG.

3.2 *Absorption spectra*

Figure 4 shows the OPA spectra along the in-plane direction of ML-BN and five BL-BN structures. For ML-BN, our calculated spectrum is consistent with previous reports [7] in terms of line shape and peak positions. To understand the spectra, we list in table 1 the transition energies and corresponding optical activities of the first two excitons of ML-BN and BL-BN. We calculated the binding energies ($E_b$) based on the direct $G_0W_0$ gap at K point because the vertical transition is considered here and the contributions to these excitons mainly come from transitions near K point [3,7]. In table 1, we also list the Davydov splitting [40,41] energy ($E_{ds}$) of BL-BN which is the energy difference between the first and second excitons. For AA-BN, it has been shown [7] that the first and second excitons mainly stem from the first exciton of ML-BN. For other four bilayer structures, we obtain similar Davydov splitting behaviors. Based on Fig. 4 and table 1, we can first see that the excitons of BL-BN exhibit large binding energies, but significantly lower than that of ML-BN, mainly due to the increased screening in the bilayer structure [7,42,43]. The first exciton of ML-BN locates at 5.25 eV and has a binding energy of 2.03 eV, in agreement with previous reports [7,44]. The binding energies of the first exciton of five bilayer structures show an order of AB-BB < AA-BB < AB-NN ≈ AB-BN

< AA-BN. In this order, the structures with the B-B opposite (*i.e.*, AB-BB and AA-BB) have a relatively small binding energy, and that those with the B-N opposite have a relatively large binding energy. Furthermore, as shown in table 1, the binding energy of the first exciton of AA-BN is 1.78 eV in agreement with the Paleari *et al.*'s report [7] in which they theoretically investigated the effects of the number of layers on the binding energies of excitonic states based on the same stacking as AA-BN. They showed that the binding energy of the first exciton of the pentalayer structure reduced to 1.32 eV which is close to those of the first excitons of AB-BB (1.38 eV) and AA-BB (1.43 eV). This implies that the electronic screening environment of the bilayer structures with the B-B opposite should be comparable to that of pentalayer AA-BN structures. Thus, the B-B opposite may have a stronger electronic screening than the B-N opposite.

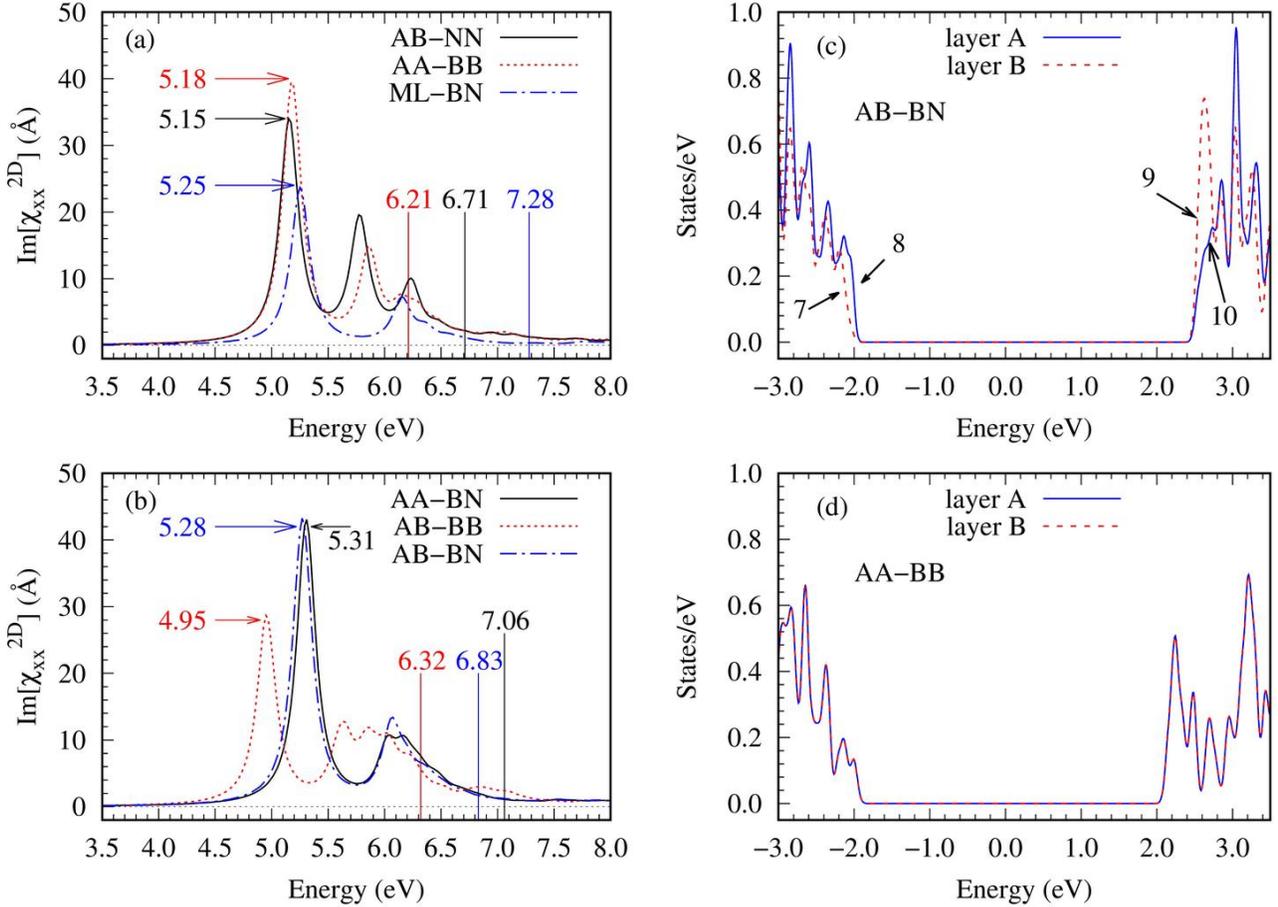

Figure 4. *(a and b) Absorption spectra (i.e., imaginary part of polarizability, Imχ) along the in-plane direction of ML-BN and BL-BN calculated using the GW+BSE method. The GW direct gaps at K point are indicated by the vertical lines. The transition energy of the first bright exciton is indicated by arrow. (c and d) PDOS per layer of AB-BN and AA-BB.*

Table 1. *Transition energies (eV) and corresponding optical activities (d = dark or br = bright) of the first two excitons of ML-BN and BL-BN. The binding energies (eV) based on the direct G0W0 gap at K point are given in parenthesis. For example, the Davydov splitting ($E_{ds}^{12}$/eV) is the energy difference between the first and second excitons. The transitions based on the IPA bands have a major contribution to the excitons.*

| Excitons | ML-BN ($D_{3h}$) | AB-BB ($D_{3d}$) | AB-NN ($D_{3d}$) | AA-BB ($D_{3h}$) | AB-BN ($C_{3v}$) | AA-BN ($D_{3d}$) |
|---|---|---|---|---|---|---|
| 1 (×2) [a] | 5.25 (2.03, br) [b] | 4.95 (1.38, d) | 5.09 (1.62, d) | 4.78 (1.43, d) | 5.24 (1.59, br) | 5.28 (1.78, d) |
| 2 (×2) | | 4.96 (1.37, br) | 5.15 (1.55, br) | 5.18 (1.03, br) | 5.28 (1.55, br) | 5.31 (1.75, br) |
| $E_{ds}^{12}$ | | 0.01 | 0.06 | 0.40 | 0.04 | 0.03 |
| 1 (×2) [a] | 4→5 (4.69) [c] | 7→9 (4.14) | 8→9 (4.47) | 8→9 (4.07) | 7→9 (4.69) | 7→9 (4.75) |
| | | 8→9 (4.14) | 8→10 (4.47) | | | 8→9 (4.72) |
| 2 (×2) | | 7→9 (4.14) | 8→9 (4.47) | 7→9 (4.45) | 8→10 (4.73) | 7→9 (4.75) |
| | | 8→9 (4.14) | 8→10 (4.47) | | | 8→9 (4.72) |

[a] "×2" means double degenerate.

[b] transition energy (binding energy, optical activity).

[c] band 4 to band 5 with the transition energy of 4.69 eV.

Second, the absorption spectra are very similar in terms of line shapes and peak positions for BL-BN with the same opposite atoms, though they have different electronic energy gaps. For example, the $G_0W_0$ gaps of AA-BN and AB-BN (Fig. 4b) are 7.06 and 6.83 eV, respectively. Both of them have strong absorption peaks at ~5.30 eV and ~6.1 eV. A similar case occurs for the AB-NN and AA-BB/NN (Fig. 4a) with the same N–N opposite atoms. We can also see that the absorption spectra of AB-BB are significantly different from those of other four bilayer structures in terms of line shapes and peak positions, especially the position of the strongest absorption peak. This absorption peak locates at 4.95 eV, which is distinctly lower than 5.20 ± 0.10 eV of other four bilayer structures. So, the AB-BB could be distinguished from other four bilayer structures by the absorption spectra.

Third, the AA-BB has the smallest $E_b$ but the largest $E_{ds}$ among five bilayer structures. To understand the large $E_{ds}$ of AA-BB, we examine the weight (*i.e.*, $|A_{c,v,k}|^2$) of contributions defined in Eq. 2 for the first and second excitons. Table 1 lists the IPA transitions at K point with the weight larger than 0.02. For example, for ML-BN, the IPA transition from 4 (HOVB) to 5 (LUCB) has a major contribution to the first bright exciton, in agreement with previous report [7]. All the bilayer structures except AA-BB have a small $E_{ds}^{12}$. For AB-BB, AB-NN, and AA-BN, the small $E_{ds}^{12}$ may be due to that two excitons stem from the same IPA transitions. For AB-BN, which also has a small $E_{ds}^{12}$ (0.04 eV), the major IPA transitions are 7→9 and 8→10 for two excitons, respectively. According to the PDOS (Fig. 4c), we find that these two transitions belong to the intralayer transition (*i.e.* 7→9 from B layer, 8→10 from A layer). Meanwhile, the difference in these two IPA transition energies is 0.04 eV (4.73 – 4.69) that is equal to the $E_{ds}^{12}$. Now, we go back to the AA-BB with the largest $E_{ds}^{12}$. The first and second excitons originate mainly from the IPA transitions of 8→9 and 7→9, respectively. According to the PDOS (Fig. 4d), we find that each layer almost has the same contribution to the transition (*i.e.* both layers contribute to the bands 7, 8, and 9). Thus, the large $E_{ds}^{12}$ is mainly due to the difference in the IPA transition channels, that is, the energy difference of 0.38 eV between bands 7 and 8 is very close to the $E_{ds}^{12}$ of 0.40 eV.

Now, we turn to TL-BN, as shown in Fig. 5, all the trilayer structures have a strong absorption peak at about 5 eV, and, to some degrees, inherit the characteristics of the absorption spectrum of the bilayer substructure. To understand these absorption peaks, we list in table 2 the information for the first three excitons of all the trilayer structures. The first three excitons are bright. Similar to BL-BN, all the first three excitons are double degenerate and related to the first exciton of each monolayer. Based on Fig. 5 and table 2, we first observe strong absorption peaks at 4.95 and 5.01 eV for NBB and BBB, respectively, which may be related to the AB-BB substructure with the lowest absorption peak at 4.95 eV (Fig. 4b). The BBN has a strong absorption peak at 5.23 eV because the substructures AA-BB and AB-BN have strong absorption peaks at 5.18 eV (Fig. 4a) and 5.28 eV (Fig. 4b), respectively. As shown in table 2, the binding energies of the first exciton of eight trilayer structures have an order of BBB (1.13) < NNB (1.21) < NNN (1.24) < NBB (1.30) < BBN (1.35) < BNB (1.52) < NBN (1.56) < BNN (1.60). In this order, we also find that the structures with the B-B opposite have a relatively small binding energy, which is obviously shown in BBB formed by AA-BB and AB-BB substructures. The substructure with the B-N opposite dramatically increases the binding energy owing to the weak electronic screening mentioned above. For example, the binding energy increases by 0.17 eV from BBN to BNB. In these two trilayer structures, the difference in the structures is AA-stacked substructure which changes from AA-BB/NN to AA-BN, and note that the AA-BN has the largest binding energy for the first two excitons (table 1).

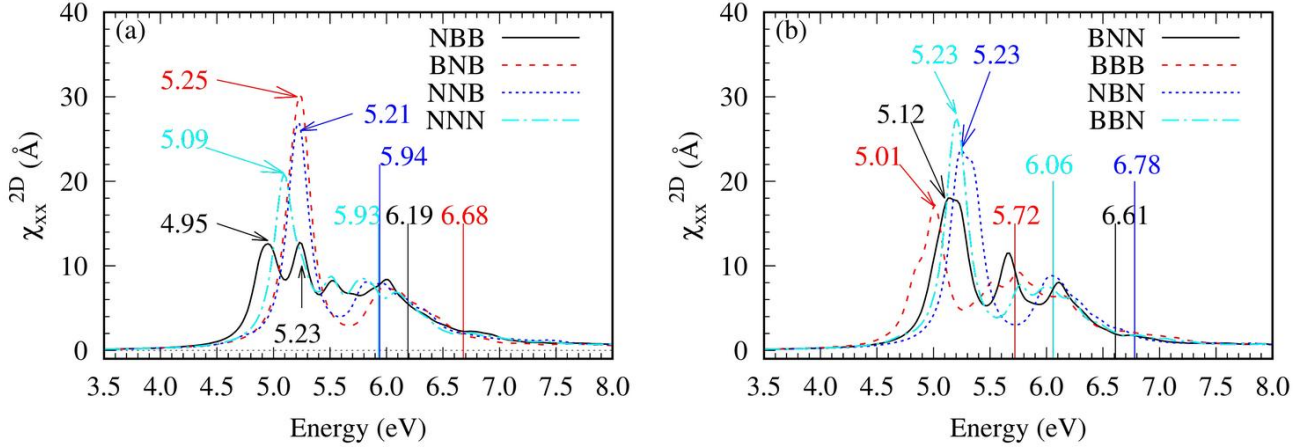

Figure 5. *Absorption spectra (i.e., imaginary part of polarizability, Imχ) along the in-plane direction of TL-BN calculated using the GW+BSE method. The GW direct gaps at K point are indicated by the vertical lines. The transition energy of the first bright exciton is indicated by arrow.*

Table 2. *Transition energies (eV) of the first three excitons (all are bright) of TL-BN. The binding energies (eV) based on the direct gap at K point are given in parenthesis. All the structures have $C_{3v}$ symmetry. The $E_{ds}^{12}$(eV) is the Davydov splitting between the first and second excitons, and the $E_{ds}^{23}$(eV) is the Davydov splitting between the second and third excitons. The transitions based on the IPA bands have a major contribution to the excitons.*

| Excitons | NNB | NNN | BBN | BBB | BNN | BNB | NBB | NBN |
|---|---|---|---|---|---|---|---|---|
| 1 (×2) [a] | 4.73 (1.21) | 4.69 (1.24) | 4.71 (1.35) | 4.59 (1.13) | 5.01 (1.60) | 5.16 (1.52) | 4.89 (1.30) | 5.22 (1.56) |
| 2 (×2) | 5.18 (0.76) | 5.04 (0.89) | 5.17 (0.89) | 4.84 (0.88) | 5.12 (1.49) | 5.21 (1.47) | 4.99 (1.20) | 5.23 (1.55) |
| 3 (×2) | 5.21 (0.73) | 5.09 (0.84) | 5.23 (0.83) | 5.01 (0.71) | 5.23 (1.38) | 5.25 (1.43) | 5.24 (0.95) | 5.36 (1.42) |
| $E_{ds}^{12}$ | 0.45 | 0.35 | 0.46 | 0.25 | 0.11 | 0.05 | 0.10 | 0.01 |
| $E_{ds}^{23}$ | 0.03 | 0.05 | 0.05 | 0.17 | 0.11 | 0.04 | 0.25 | 0.13 |
| 1 (×2) [a] | 12→13(4.05) | 12→13(3.97) | 12→13(4.05) | 12→13(3.79) | 12→13(4.48) | 11→14(4.73) | 12→13(4.14) | 10→13(4.70) 12→14(4.75) |
| 2 (×2) | 10→13(4.44) | 12→14(4.40) | 10→13(4.45) | 11→13(3.98) | 12→14(4.48) | 12→15(4.75) | 10→13(4.19) | 10→13(4.70) 12→14(4.75) |
| 3 (×2) | 11→13(4.14) | 11→13(4.24) | 11→14(4.70) | 10→13(4.19) | 11→15(4.75) | 10→13(4.74) | 10→14(4.76) | 11→15(4.78) |

[a] "×2" means double degenerate.

Second, the magnitude of two Davydov splitting energies ($E_{ds}^{12}$ and $E_{ds}^{23}$) of the trilayer structures is closely related to that of $E_{ds}^{12}$ of the bilayer substructures. As shown in table 1, the $E_{ds}^{12}$ of five bilayer structures can be ordered by AA-BB (0.40) > AB-NN (0.06) > AB-BN (0.04) ≈ AA-BN (0.03) > AB-BB (0.01). From table 2, we can see that the trilayer structures (NNB and BBN), containing the AA-BB/NN and AB-BN substructures, have relatively large $E_{ds}^{12}$ but relatively small $E_{ds}^{23}$. For NNB and BBN, the $E_{ds}^{12}$ is 0.45 and 0.46 eV, respectively, which is even larger than that of AA-BB (0.40 eV in table 1). This is due to a large $E_{ds}^{12}$ (0.40 eV) of AA-BB/NN and a small one (0.04 eV) of AB-BN. A similar case occurs for BBN which has the $E_{ds}^{12}$ and $E_{ds}^{23}$ of 0.35 and 0.05 eV, respectively. Both BNB and NBN have relatively small $E_{ds}^{12}$ and $E_{ds}^{23}$ because they contain the AA-BN and AB-BN substructures with relatively small $E_{ds}^{12}$ (*i.e.* 0.03 and 0.04, respectively). Interestingly, for BBB containing the AA-BB/NN and AB-BB substructures, the AA-BB substructure has the largest $E_{ds}^{12}$ (0.40 eV) among the five bilayer structures while the AB-BB substructure has the smallest $E_{ds}^{12}$ (0.01 eV), which leads to similar values of $E_{ds}^{12}$ (0.25 eV) and $E_{ds}^{23}$ (0.17 eV).

Finally, according to the IP transition contributions to excitons, we can see that the $E_{ds}$ values of excitons of TL-BN also depend on the IP transition energies strongly, similar to those of BL-BN. For example, a large $E_{ds}^{12}$ (0.45 eV) of NNB is related to a large difference in transition energies between 10→13 (4.44 eV) and 12→13 (4.05 eV) transitions. The $E_{ds}^{12}$ and $E_{ds}^{23}$ of BNB have similar values (0.05 and 0.04 eV, respectively) because the IP transitions with major contributions have very similar transition energies (~ 4.74 eV).

### 3.3 *Direct FBG based on the B–B opposite substructure*

As shown above, the B–B opposite makes the multilayer BN structures trend to having a direct FBG (Figs. 2c, 2e, 2g, 2h, and 2i). Particularly, for AB-BB (Fig. 2e) and AAB-BBB (Fig. 2h), which have as many B–B opposites as possible, the FBG is direct within both the IPA and GWA. However, the N–N opposite results in an indirect FBG gap (Figs. 2f, 2k, and 2l). In this section, we explore the effect of the N–N opposite on the FBG of the B–B opposite structures to show that the reservation of the B–B opposite substructure plays an important role in forming the direct FBG. For this purpose, we construct a series of six-layer structures by inserting the N–N opposite into the six-layer AB-BB structure and keep at least a B–B opposite in these AB-stacked structures. These structures are labeled by the order of the opposite atoms, similar to those shown in Fig. 1. As an example, the geometry of BNNBBB is given in Fig. 6k. We calculated their energy band structures within the IPA, and the results are shown in Figs. 6(a–j).

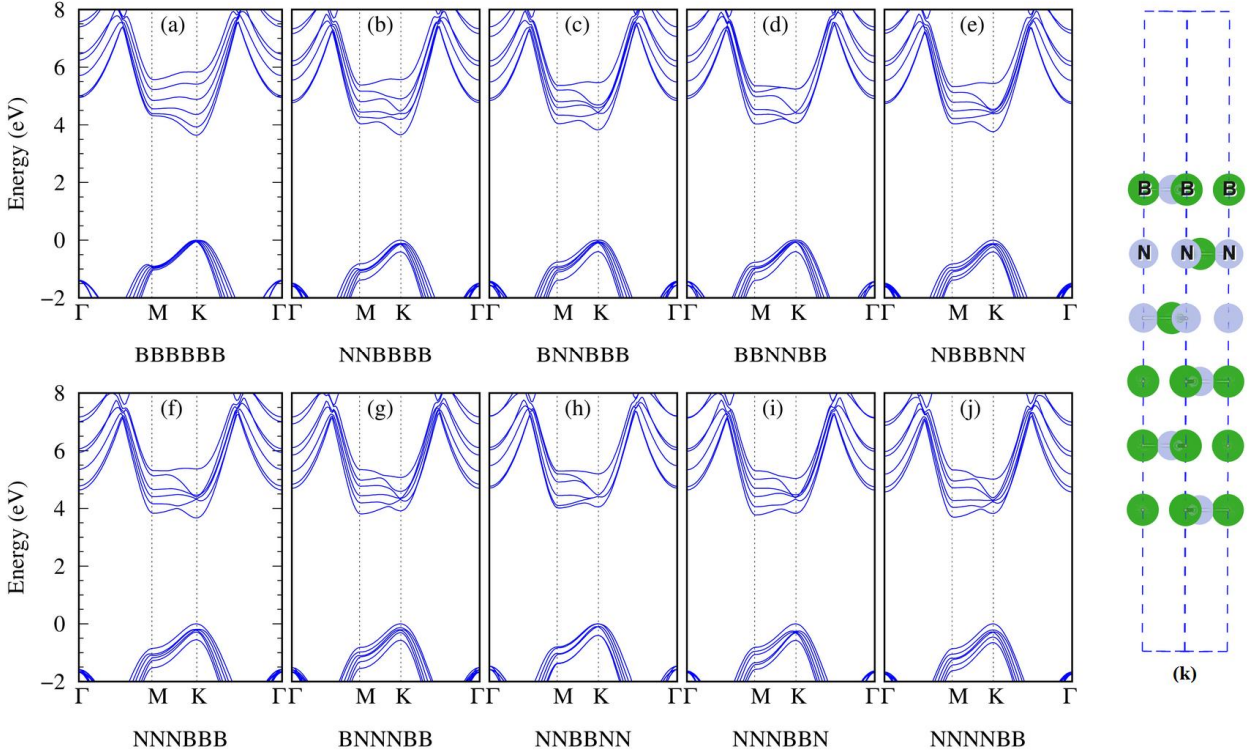

Figure 6. *(a–j) Band structures of six-layer BN structures based on the IPA calculations. The labels (e.g., NNBBBB and BNNBBB) are defined by the directly opposite atoms. (k) Geometry of BNNBBB.*

As shown in Fig. 6a, the six-layer AB-stacked structure with B–B opposite (BBBBBB) has a direct FBG of 3.64 eV at K point. By inserting N–N opposite into the BBBBBB, the structure gradually transforms from direct FBG (Figs. 6a–6c, 6e, and 6f) to indirect one (Figs. 6g–6j) as the number of N–N opposites increases. The structures with one N–N opposite at different insertion positions have a direct FGB (Figs. 6b, 6c, and 6e) because two or three B–B opposites are reserved in the structure. A consecutive B–B opposite unit is essential for the structure to have a direct FGB, as shown in Figs, 6b, 6c, 6e, and 6f whose corresponding structures have BBB unit. On the contrary, consecutive N–N opposites (Figs. 6g and 6i) make the structure exhibit an indirect FBG. To understand these behaviors, we show in Fig.7 the PDOS of several selected six-layers. The conduction band edges of BNNNBB (Fig. 7d) and NBBBNN (Fig. 7e) mainly come from NNN and BBB units, respectively, which leads to indirect and direct FBG in corresponding structures. For NNNBBB (Fig. 7b), NNN and BBB competitively contribute to the conduction band edge, and the result is that NNNBBB exhibit a direct FBG determined by the BBB unit.

Meanwhile, we can see that the inner layers have more effects on the band edge than the outer layers,

because the inner layers interact with two sides, which possibly leads to a larger dispersion of energy band. For example, in both BBNNBB and NNBBNN, the B–N opposites of inner layers play a major role in determining the type of FBG. As shown in Figs. 2b, 2d, and 3a, the bilayer structures with the B–N opposite have a relative flat M–K path and a slightly indirect FGB, which leads to a slightly indirect FBG in both BBNNBB (Fig. 6d) and NNBBNN (Fig. 6h). Similarly, for BNNNBB (Fig. 7d) and NBBBNN (Fig. 7e), the inner NNN and BBB units also determine the type of FBG (the former is indirect and the latter is direct). Interestingly, our present finding is very applicable to the ten-layer t-BN reported by Mengle and Kioupakis [23]. Although they only show a representative t-BN structure (Fig. 3a of Ref[23]) consisting of ten randomly chosen layers, we can see from this structure that two inner BBB units lead the structure to having a direct FBG.

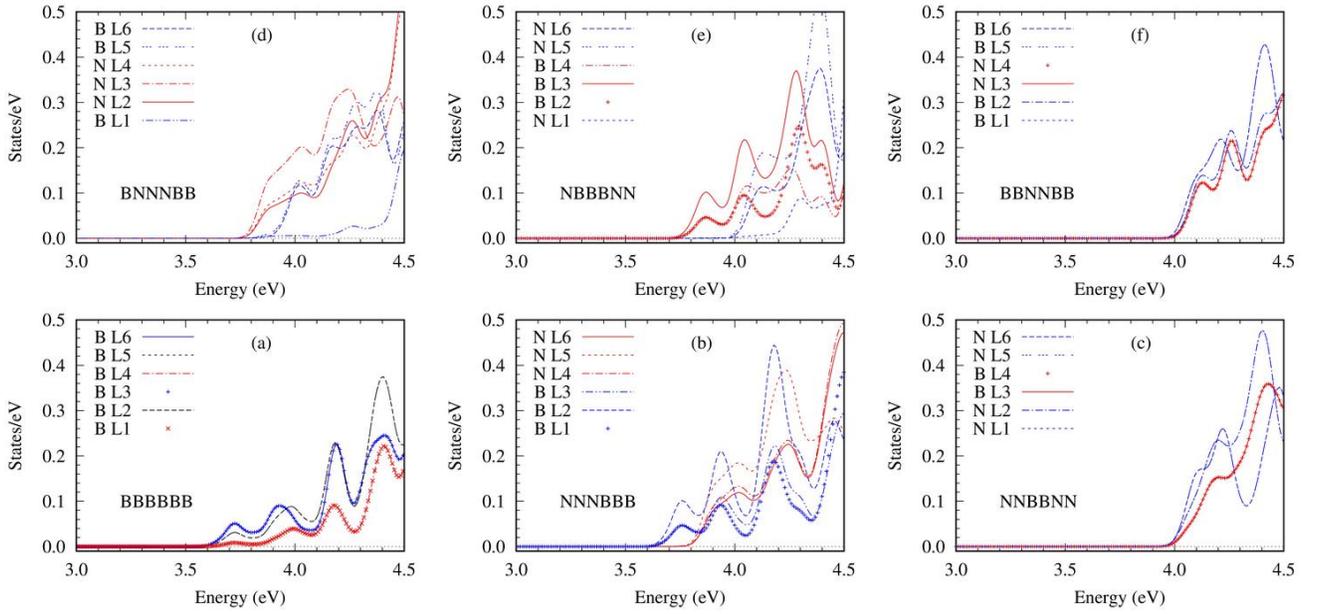

Figure 7. *PDOS of selected six structures based on the $p_z$ orbitals of B and N atoms. L1 represents the first layer, L2 represents the second layer, and so on. The atomic labels in the legends indicate the directly opposite atoms in the six-layer structure.*

More examples, we show in Fig. S2 the band structures of selected AB-stacked four- and five-layer structures. As expected, the BBBNN and NBBBN with a BBB unit have a direct FGB, while the BBNNN and BNNNB with a NNN unit have an indirect one. The NBBNN with not only a inner B–B opposite but also B–N and N–N opposites exhibit a relative flat M-K path and have a slightly direct FBG at K point, a similar case occurs for BNNBB which has a slightly indirect FBG. The BBNN with not only B–B opposite but also N–N or N–B opposite has a slightly direct FBG at K point. Finally, BBNBB has an obvious direct FBG at K points which is determined by two B–B opposites, though two N–B opposites locate in inner layer. Thus, the B–B opposite plays a crucial role in determining the type of FBG.

## 4. Conclusions

We have performed the first-principles many-body GW and BSE calculations on the BL-BN and AAB-stacked TL-BN structures. The size and type of FBG strongly depend on the opposite atoms. Structures dominated by the B–B opposite are expected to have a direct FBG. The B–B opposite makes the stacking have a relative small binding energy of exciton, and thus the corresponding structures have a stronger electronic screening than those with the B–N and N–N opposites. The Davydov splitting energies of excitons of TL-BN are closely related to those of BL-BN substructure, which implies the Davydov splittings of FL-BN could be understood on the basis of those of substructure. All the structures have similar dispersion of valence band edge, and thus the intrinsic FBG is mainly determined by the conduction band edge whose dispersion depends on the type of opposite atoms in FL-BN. For $t$-BN or FL-BN, to obtain the structure with a direct FBG, we should construct the B–B opposite as many as possible and preferably locate them in the inner layer because the B–B opposite can make the structure have a direct FBG at K point. Our findings not only show a new structure-property relationship but also provide a useful reference for experimentally designing FL-BNs with a direct FBG which are more suitable for optoelectronic applications in deep ultraviolet region.

**Acknowledgements**

We appreciate the financial support from Natural Science Foundation of China Project 21303164.

# Effects of opposite atoms on electronic structure and optical absorption of two-dimensional hexagonal boron nitride

You-Zhao Lan*[1]

*Key Laboratory of the Ministry of Education for Advanced Catalysis Materials, College of Chemistry and Life Sciences, Zhejiang Normal University, Jinhua, Zhejiang, 321004, China*

Table S1. Convergence tests of the energy gap of ML-hBN based on the change of the number of empty states, response block size, and k-grid. 1 Ry = 13.6 eV.

|  | Number of empty states : 200; response block size : 5 Ry | | | |
|---|---|---|---|---|
| k-grids | 18×18×1 | 24×24×1 | 30×30×1 | 36×36×1 |
| Gap at K (eV) | 7.32 | 7.27 | 7.24 | 7.24 |
| Global gap (eV) | 6.63 | 6.49 | 6.44 | 6.41 |
|  | k-grid: 30×30×1; response block size : 5 Ry | | | |
| Number of empty states | 100 | 200 | 300 | 400 |
| Gap at K (eV) | 7.25 | 7.24 | 7.24 | 7.25 |
| Global gap (eV) | 6.29 | 6.41 | 6.43 | 6.45 |
|  | Number of empty states : 200; k-grid : 30×30×1 | | | |
| Response block size (Ry) | 5 | 8 | 10 | 13 |
| Gap at K (eV) | 7.24 | 7.25 | 7.28 | 7.29 |
| Global gap (eV) | 6.41 | 6.52 | 6.57 | 6.59 |

---

[1] Corresponding author: lyzhao@zjnu.cn

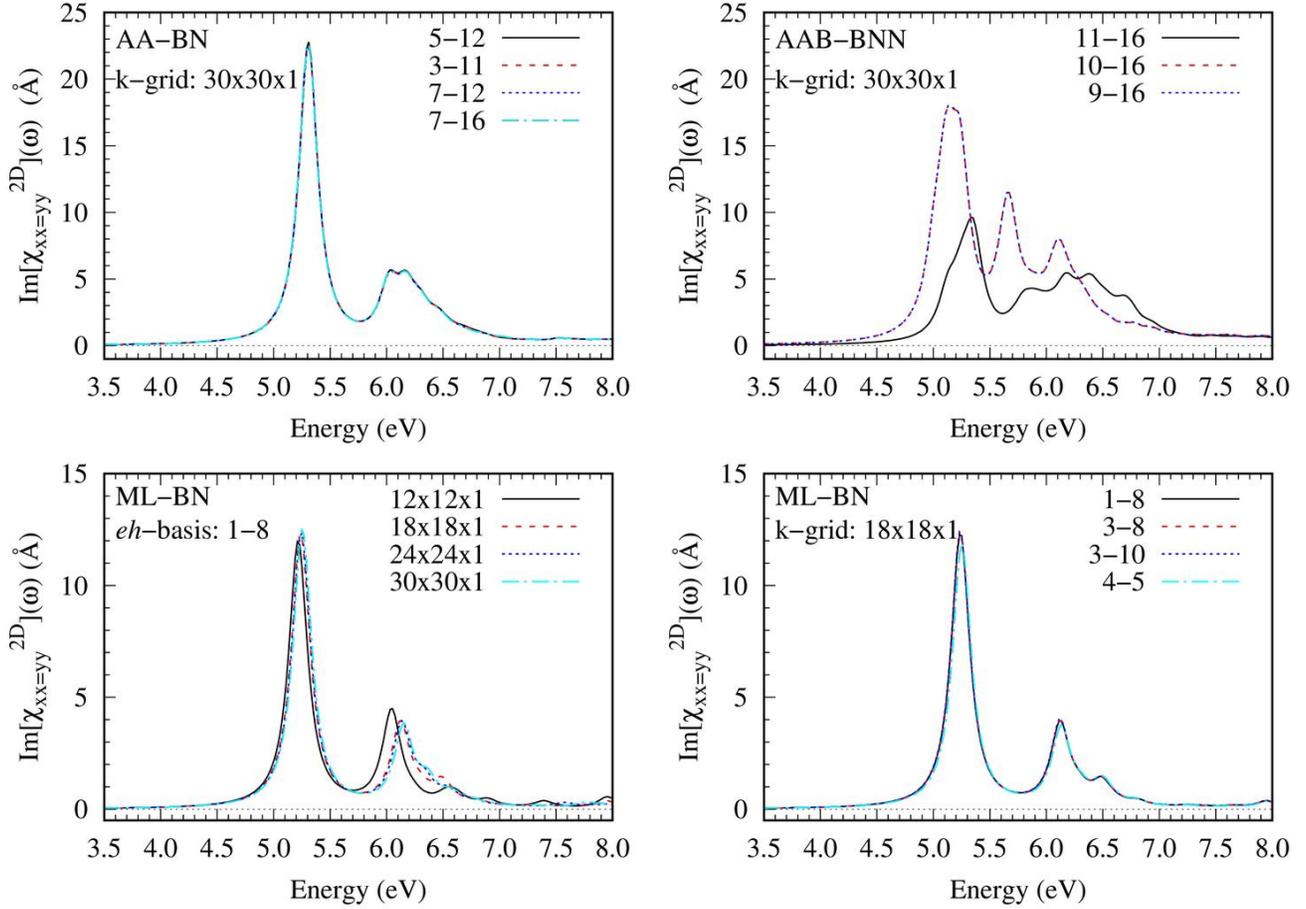

Figure S1. The dependence on the k-grid and *eh*-basis of the imaginary part of two-dimensional polarizability (Im$\chi^{2D}$) of ML-BN, AA-BN, and AAB-BNN. The legend of "5–12" is defined on the basis of the index of bands of IPA. For example, in the case of AA-BN with 16 valence electrons occupying 8 valence bands, it indicates that the highest four valence bands and the lowest four conduction bands are included in the BSE calculation.

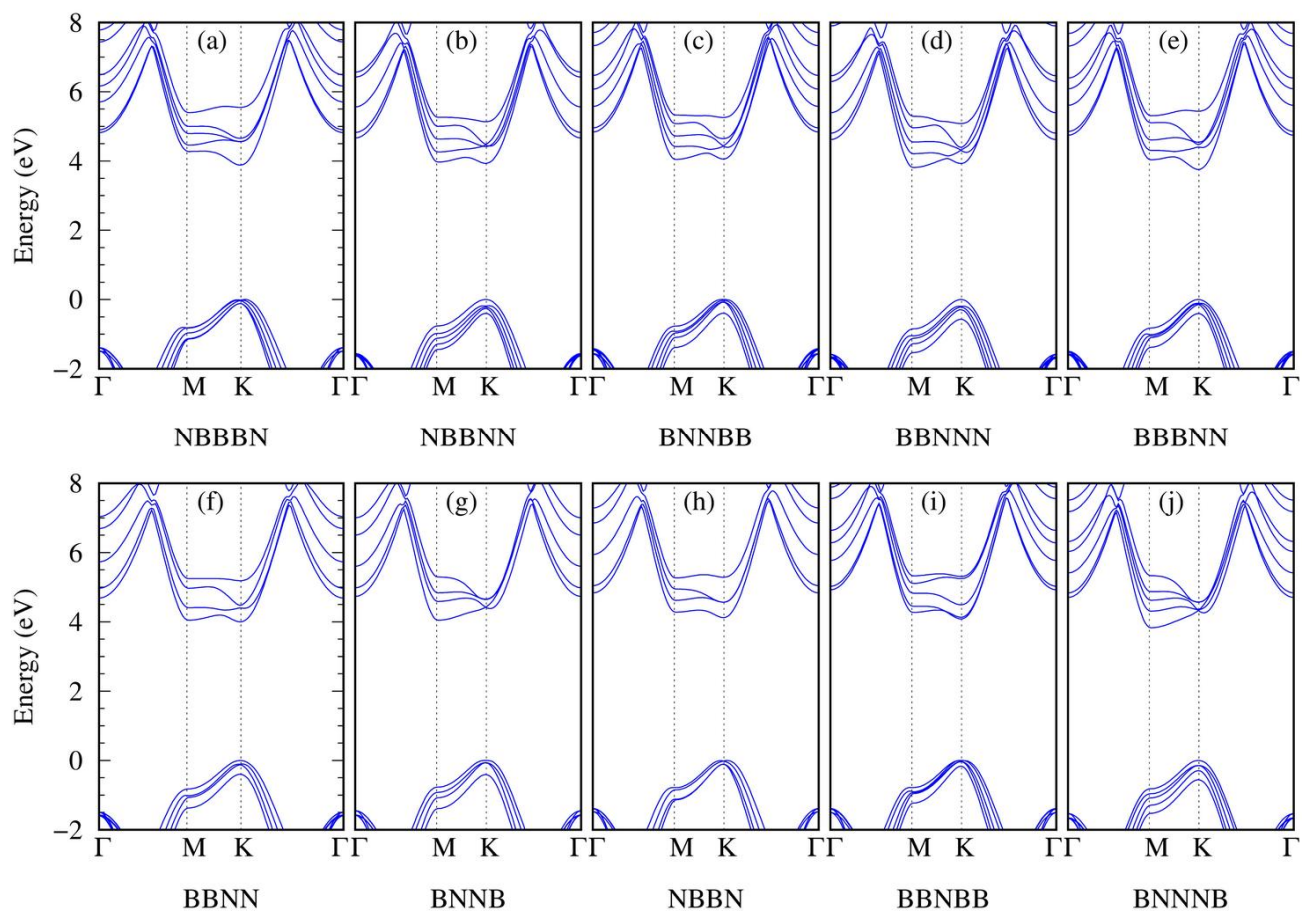

Figure S2. Band structures of selected four- and five-layer AB-stacked structures.